\documentclass[10pt]{article}
\usepackage{amsthm}
\usepackage{amsmath,amssymb,enumerate}
\title{\large Conditional and Unique Coloring of Graphs\\  \medskip }
 \author{        
 \small P. Venkata Subba Reddy and K. Viswanathan Iyer \thanks{author for correspondence}   \\
 \small Dept. of Computer Science and Engg. \\
 \small National Institute of Technology \\
 \small Tiruchirapalli - 620015, India \\
 \small E-mail: venkatpalagiri@gmail.com, vichuiyer@gmail.com \\ }
 \date{}
\begin{document}
\maketitle
\begin{abstract}
For integers $k, r > 0$, a conditional $(k,r)$-coloring of a graph $G$ is a proper $k$-coloring of the vertices of $G$ such that every vertex $v$ of degree $d(v)$ in $G$ is adjacent to at least $\min\{r, d(v)\}$ differently colored vertices. Given $r$, the smallest integer $k$ for which $G$ has a conditional $(k,r)$-coloring is called the $r$th order conditional chromatic number $\chi_r(G)$ of $G$. We give results (exact values or bounds for $\chi_r(G)$, depending on $r$) related to the conditional coloring of some graphs. We introduce \emph{unique conditional colorability} and give some related results.\\ \\*
 \bigskip 
\end{abstract} 
\section{Introduction}
Let $G= (V(G),E(G))$ be a simple, connected, undirected graph. For a vertex $v \in V(G)$, the \textit{neighborhood} of $v$ in $G$ is defined as $N_G(v)$= \{$u \in V(G):(u,v) \in E(G)$\}, and the degree of $v$ is denoted by $d(v)$=$|N_G(v)|$. The \textit{closed neighborhood} of $v$ is defined as $N_G[v] = N_G(v) \cup \{v\}$. For an integer $k>0$, a \textit{proper} $k$-\textit{coloring} of a graph $G$ is a surjective mapping $c \colon V(G) \to \{1,2,\ldots,k \}$ such that if $(u,v) \in E(G)$ then $c(u) \neq c(v)$. The smallest $k$ such that $G$ has a proper $k$-coloring is the \textit{chromatic number} $\chi(G)$ of $G$. For a set $S \subseteq V(G)$ we define $c(S)= \{c(u) : u \in S \}$. \medskip \\ 
\noindent \textbf{Definition 1} For integers $k,r > 0$, a \textit{conditional} $(k,r)$-\textit{coloring} of  $G$ is a surjective mapping $c \colon V(G) \to \{1,2,\ldots,k \}$ such that both the following conditions (C1) and (C2) hold:
\begin{quote}
(C1) if $(u,v) \in E(G)$, then $c(u) \neq c(v)$. \\
(C2) for any $v \in V(G)$, $|c(N_G(v))| \geq $ min \{$d(v),r$\}. 
\end{quote}
Given an integer $r>0$, the smallest integer $k$ such that $G$ has a proper $(k,r)$-coloring is called the \textit{rth-order conditional chromatic number} of $G$, denoted by $\chi_r(G)$.
It is proved in \cite{Li09} that the conditional $(k, r)$-coloring problem of a graph is $NP$-complete. Two of our earlier work can be found in \cite{reddy1,reddy2}. We follow commonly used terminology and notations (see for example \cite{Gol80,Har69,West03}).
\section{Unique conditional colorability}
If $\chi(G)=k$ and every $k$-coloring of $G$ induces the same partition of $V(G)$, then $G$ is called \textit{uniquely k-colorable}. In a similar way we define unique $(k,r)$-colorability of graphs. \medskip \\ 
\noindent \textbf{Definition 2} If $\chi_r(G)=k$ and every conditional $(k,r)$-coloring of $G$ induces the same partition of $V(G)$, then $G$ will be called \textit{uniquely} $(k,r)$-\emph{colorable}.
\newtheorem{pro1}{Proposition}  
\begin{pro1}
If $G$ is uniquely $n$-colorable and $r \leq n-1$ then $\chi_r(G) = n$. 
\end{pro1}
\begin{proof}
Since $G$ is uniquely $n$-colorable, let $c \colon V(G) \to \{1,2,\ldots,n \}$ be the proper coloring of $G$ and w.l.o.g. for $1\leq i \leq n$ let the color class $C_i$ be defined as $C_i=\{v : c(v)=i \}$. For all $u \in V(G)$ if $u \in C_i$ then for all $j \in \{1,2,\ldots,n \}$ there exists a $v \in C_j \;(j \neq i)$ -- this implies that for every $u \in V(G)$, $d(u) \geq n-1 $ and $|c(N(u))|= n-1$. Note that $c$ is also a conditional $(n,r)$-coloring of $G$ because (C2) is also satisfied as for every $u \in V(G),|c(N(u))|= n-1 \geq $ \ min $\{r,d(u)\}$.   
\end{proof}
The definition of conditional $(k,r)$-coloring of $G$ and Proposition 2.1 together imply:
\newtheorem{cor}{Corollary}  
\begin{cor}
Every uniquely $n$-colorable graph $G$ is also uniquely $(n,n-1)$-colorable.
\end{cor}
\newtheorem{pro2}[pro1]{Proposition}  
\begin{pro2}
For every $k \geq 3$, there exists a uniquely $(3,2)$-colorable graph $G_k$ with $k+2$ vertices.
\end{pro2}
\begin{proof}
We take $G_1$ to be $C_3$. Suppose that $k \geq 1$ and assume  that $G_k$ has been obtained. From $G_k$, we construct $G_{k+1}$ by introducing a new vertex $w$. The vertex and edge sets of $G_{k+1}$ are defined thus:
\begin{align*}
V(G_{k+1}) &= V(G_k) \cup \{w\}, \text{where}\; w \notin V(G_k). \\
E(G_{k+1}) &= E(G_k) \cup \{(u,w),(v,w)\}, \text{where} \ (u,v) \in E(G_k). 
\end{align*}
Evidently,  $|V(G_k)| = k+2$. We need to show that for any $k$, $G_{k}$ is uniquely $(3,2)$-colorable. We prove the result by induction on  $k$. Clearly $G_1$ is uniquely $(3,2)$-colorable. For some $k>1$, assume $G_k$ is uniquely $(3,2)$-colorable. Then $G_{k+1}$ is also uniquely $(3,2)$-colorable because the new vertex $w \in G_{k+1}$ is assigned a third color different from its two neighbors which are adjacent; by the inductive hypothesis the result follows.
\end{proof}
\newtheorem{pro3}[pro1]{Proposition}  
\begin{pro3}
Every path $P_n \;(n \geq 3)$ is uniquely $(3,2)$-colorable.
\end{pro3}
\begin{proof}
From \cite{Lai06} we get $\chi_2(P_n)= 3$ . Define a conditional $(3,2)$-coloring $c \colon V(P_n) \to \{1,2,3 \}$ by    
 \begin{align*}
  c^{-1}(1) = C_1  &=  \{v_i : i \bmod 3= 1 \}, \\
  c^{-1}(2) = C_2  & = \{v_i : i \bmod 3= 2 \}, \\     
  c^{-1}(3) = C_3  & = \{v_i : i \bmod 3= 0 \},
 \end{align*} 
where, $C_1$,$C_2$ and $C_3$ are the color classes. In the conditional $(3,2)$-coloring of $P_n$, for any two vertices  $v_i,v_j \; (i \ne j)$ in $V(P_n)$ if $|i-j| \bmod 3 = 0$ then $v_i$ and $v_j$ must be colored same; otherwise either (C1) will be violated at $v_{min\{i,j\}+1}$ or (C2) will be violated at $v_{max\{i,j\}-1}$. Since $c$ is the only coloring wherein for all $v_i,v_j \in V(P_n)$, $c(v_i) = c(v_j)$, if $|i-j| \bmod 3= 0$,  $P_n$ is uniquely $(3,2)$-colorable.    
\end{proof}
\newtheorem{pro4}[pro1]{Proposition}  
\begin{pro4}
If $T$ ($\neq P_n$) is a rooted tree with $n$ verices and $k = \chi_r(T)$ then $T$ is not uniquely $(k,r)$-colorable unless $k=n$. 
\end{pro4}
\begin{proof}
Let the root of $T$ be a vertex of degree $\Delta(T)$. Let $c \colon V(T) \to \{1,2,\ldots,k \}$ be a conditional $(k,r)$-coloring of $T$. We show that a new conditional $(k,r)$-coloring of $T$ can be obtained based on $c$ if $k \neq n$. We know that every tree $T$ has  at least two vertices say, $u,v$ with degree less than two. Let $p(v)$ deonte the parent of $v$. We have the following cases:\\ 
\textbf{Case 1:} $p(u)= p(v):$ \\ 
\textbf{Case 1.1:} $c(u) = c(v):$ Since $r \geq 2$ assigning one of the colors from the set  $c(N_T(p(u))) \setminus \{c(v)\}$ to $u$ results in a new conditional $(k,r)$-coloring. \\
\textbf{Case 1.2:} $c(u) \neq c(v):$ \\
\textbf{Case 1.2.1:} There exists a vertex $w \in V(T)$ such that $c(w) = c(u)$ or $c(w)=c(v):$ Interchanging the colors of $u$ and $v$ results in a different induced partition of $V(T)$.\\
\textbf{Case 1.2.2:} There doesn't exist a vertex $w \in V(T)$ such that $c(w) = c(u)$ or $c(w)=c(v):$\\
\textbf{Case 1.2.2.1:} $r \geq \Delta :$ If $n \neq \Delta+1$ then there exists a vertex $w' \in N_T(p(u))$ such that $d(w') \geq 2$; then interchanging the colors of $u$ and $w'$ results in a different induced partition of $V(T)$ because the subtree rooted at $w'$ doesn't contain any vertex colored $c(u)$. If $n = \Delta+1$ then $k=n$.\\
\textbf{Case 1.2.2.2:} $r < \Delta :$ There must exist at least two vertices $w_1, w_2 \in N_T(p(u))$ such that $c(w_1) = c(w_2)$. Interchanging the colors of $u$ and $w_1$ results in a different induced partition of $V(T)$ because the subtree rooted at $w_1$ doesn't contain any vertex colored $c(u)$. \\
\textbf{Case 2:} $p(u)\neq  p(v):$  \\
\textbf{Case 2.1:} $d(p(u)) < $ min $\{r,\Delta(T)\}$ or $d(p(v)) < $ min $\{r,\Delta(T)\} :$ Assigning to $u$ any color in the set $c(V(T)) \setminus c(N_T[p(u)])$ if $d(p(u)) < $ min $\{r,\Delta(T)\}$ or to $v$ any color in the set $c(V(T)) \setminus c(N_T[p(v)])$ if $d(p(v)) < $ min $\{r,\Delta(T)\}$ gives a new conditional $(k,r)$-coloring. \\
\textbf{Case 2.2:}  $d(p(u)) > $ min $\{r,\Delta(T)\}$ or $d(p(v)) > $ min $\{r,\Delta(T)\} :$ If $d(p(u)) > $ min $\{r,\Delta(T)\}$ there must exist a vertex $u' \in N_T(p(u))$ such that $c(u) = c(u')$ or two vertices $u_1,u_2 \in N_T(p(u))$ such that $c(u_1) = c(u_2) \neq c(u)$. Assigning to $u$ any of the color in the set $c(V(T)) \setminus \{c(u'),c(p(u))\}$ or  intechanging the colors of $u$ and $u_1$ and the colors $c(u)$, $c(u_1)$ in the subtree rooted at $u_1$ gives a new conditional $(k,r)$-coloring in the former and latter cases respectively. The case $d(p(v)) > $ min $\{r,\Delta(T)\}$ is similar.\\ 
\textbf{Case 2.3:}  $d(p(u)) = d(p(v)) = $ min $\{r,\Delta(T)\} :$ \\
\textbf{Case 2.3.1:} min $\{r,\Delta(T)\} > 2 :$ There exists a $w \in V(T)$ such that $c(u) \neq c(w)$ and $p(u)=  p(w)$. Making the color of $u$ as $c(w)$ and in the subtree rooted at $w$,  swapping the colors $c(u)$ and $c(w)$ gives a new conditional $(k,r)$-coloring. \\
\textbf{Case 2.3.2:} min $\{r,\Delta(T)\} = 2 :$ Since $T \neq P_n$ so $\Delta(T) > 2$ and $r=2$. There exists an ancestor of $u$ and $v$ with degree $\geq 2 $ because the root of $T$ has maximum degree. Let $w$ with $d(w) \geq 2$ be the closest ancestor of $u$ and $v$. Then there exists two children of $w$ namely $w_1$ and $w_2$ which are ancestors of $u$, $v$ respectively. If $w$ is the root of $T$ and $c(w_1) \neq c(w_2)$, interchanging the colors $c(w_1)$ and $c(w_2)$ in the subtree rooted at $w_1$ gives a new conditional $(k,r)$-coloring. If $w$ is not the root of $T$ and $c(w_1) \neq c(w_2)$, interchanging the colors $c(w_1)$ and $c(w_2)$ in the subtree rooted at $w$ gives a new conditional $(k,r)$-coloring. Otherwise (i.e., if $c(w_1) = c(w_2)$), there exists a $w_3 \in N_T(w)$ such that $c(w_3) \neq c(w_1)$ and interchanging the colors $c(w_1)$ and $c(w_2)$ in the subtree rooted at $w_1$ gives a new conditional $(k,r)$-coloring.
\end{proof}
\section{Conditional colorability of some graphs}	
\newtheorem{thm1}{Theorem}  
\begin{thm1}
Let $G_1$ and $G_2$ be two graphs where $\chi(G_1)=k_1$, $\chi(G_2)=k_2$ and w.l.o.g. let $k_1 \le k_2$. Then $\chi_r (G_1 + G_2)$ = $\chi (G_1+ G_2)$ = $k_1 + k_2$, where  $r \leq k_1+ 1$.
\end{thm1}
\begin{proof}
In  the graph $G_1+ G_2$, $V(G_2) \subset N_{G_1+ G_2}(u)$ if $u \in V(G_1)$ or $V(G_1) \subset N_{G_1+ G_2}(u)$ if  $u \in V(G_2)$. Therefore $c(V(G_1)) \cap c(V(G_2)) = \emptyset $, and in any proper $k$-coloring of $G_1+ G_2$, for all $u \in V(G_1+ G_2)$, $|c(N_{G_1+G_2}(u))| \geq \min \{d(u),r \} $. This implies that  every proper $k$-coloring of $G_1+ G_2$ is also a proper $(k,r)$-coloring -- if not, we get a contradiction:  suppose that $\chi(G_1 + G_2) = k \neq k_1 + k_2$; since $c(V(G_1)) \cap c(V(G_2)) = \emptyset $, either $k_1 \neq \chi(G_1)$  or  $k_2 \neq \chi(G_2)$ which contradicts the given condition.  
\end{proof}
\newtheorem{thm2}[thm1]{Theorem}  
\begin{thm2}
Let $G(V_1,V_2,E)$ be a bipartite graph, $S_1=\bigcap_{u \in V_1}N_G(u)$, $S_2=\bigcap_{v \in V_2}N_G(v)$ and w.l.o.g. let $|S_1| \le |S_2|$. Then  $\chi_r(G) = 2r$ where $r \leq |S_1|$ . 
\end{thm2} 
\begin{proof}
In any proper coloring of $G$, from the given conditions $|c(V_1)| \geq r$ and $|c(V_2)| \geq r$ as $G$ is bipartite. Since $r \leq |S_1|$ and $c(S_1) \cap c(S_2) = \emptyset $ we have $\chi_r(G) \geq 2r$. But there exists a proper $2r$-coloring of $G$ such that $|c(S_1)|= |c(S_2)|=r$ because every bipartite graph is bicolorable and $r \geq 2$. This coloring also satisfies (C2) as $S_1 \subseteq V_2$ and $S_2 \subseteq V_1$. Thus  $\chi_r(G) \leq 2r$. Hence $\chi_r(G) = 2r$.
\end{proof}
\newtheorem{thm3}[thm1]{Theorem}  
\begin{thm3}
Let $T_1, T_2$ be two non trivial trees with $n_1,n_2$ number of vertices respectively and w.l.o.g. let $n_1 \leq n_2$. Then $\chi_r(T_1+T_2)= 2(r-1)$, where $4 \leq r \leq \  n_1+1$. 
\end{thm3}
\begin{proof}
Every nontrivial tree has at least two vertices with degree one \cite{Har69}. Therefore there exist vertices $u,v$ where $u \in V(T_1)$ and $v \in V(T_2)$, such that $d(u)=d(v)=1$. Therefore, $d_u(T_1+T_2)= 1+n_2$ and $d_v(T_1+T_2)= 1+n_1$. If $\chi_r(T_1+T_2) < 2(r-1)$, then either $|c(V(T_1))| < r-1$ or $|c(V(T_2))| < r-1$ or both because $c(V(T_1)) \cap c(V(T_2)) = \emptyset $. Hence (C2) is violated at $u$ or $v$ or both. Therefore, $\chi_r(T_1+T_2) \geq 2(r-1)$. Since every tree is $2$-colorable and $r \geq 4 $, properly color $V(T_1)$, $V(T_2)$ in $T_1+T_2$ using $r-1$ colors each such that $|c(V(T_1+T_2))|=2(r-1)$. The resulting coloring is  a conditional $(2(r-1),r)$-coloring of $T_1+T_2$, as (C1) is satisfied because $c(V(T_1)) \cap c(V(T_2)) = \emptyset $, and (C2) is satisfied because $c(V(T_1)) \cap \  c(V(T_2)) = \emptyset$ and for all $w \in V(T_1+T_2)$, $|c(N_{T_1+T_2}(w))| \geq (r-1)+1 \geq$  min  $\{d(w),r \}$. Hence the result. 
\end{proof}
\newtheorem{thm4}[thm1]{Theorem}  
\begin{thm4}
Given any two graphs $G_1$ and $G_2$, let $r_1$ and $r_2$ be such that $r_1 \geq \delta(G_1)$ and $r_2 \geq \delta(G_2)$.  Then  $\chi_r(G_1 \ \Box \ G_2) \leq \chi_{r_1}(G_1). \chi_{r_2}(G_2)$  where $r \leq \delta(G_1)+ \delta(G_2)$.
\end{thm4}
\begin{proof}
Let $\chi_{r_1}(G_1)= g_1$ and $\chi_{r_2}(G_2)= g_2$. Let $c_{G_1}$ (resp. $c_{G_2}$) be a proper $(g_1,r_1)$- (resp. $(g_2,r_2)$-) coloring of $G_1$ (resp. $G_2$). Then let $c_{G_1 \Box G_2}$ be a coloring of $G_1 \Box G_2$ wherein we assign to any vertex $(u_1,u_2) \in V(G_1 \ \Box \ G_2)$ the color denoted by the ordered pair $(c_{g_1}(u_1),c_{g_2}(u_2))$. This coloring uses $g_1.g_2$ colors and it defines a proper coloring of $G_1 \ \Box \ G_2$. Therefore $c_{G_1 \Box G_2}$ satisfies (C1).  
Let $(u_1,u_2) \in V(G_1 \ \Box \ G_2)$ such that $u_1 \in V(G_1)$ and $u_2 \in V(G_2)$. Since $c_{G_1}$ and $c_{G_2}$ satisfy (C2), by the definition of $G_1 \Box G_2$, a vertex $(u_1,u_2)$ has at least $\min \{r_1, \delta(G_1)\}=\delta(G_1)$ distinctly colored neighbors of the form $(u',u_2)$ because $|c(N_{G_1}(u_1))| \geq  \delta(G_1)$ and at least $\min \{r_2, \delta(G_2)\}=\delta(G_2)$ distinctly colored neighbors of the form $(u_1,u'')$  because  $|c(N_{G_2}(u_2))| \geq  \delta(G_2)$. Therefore $|c(N_{G_1 \ \Box \ G_2}((u_1,u_2))| \geq \delta(G_1)+ \delta(G_2)\geq r$. Hence $c_{G_1 \Box G_2}$ satisfies (C2) and the result follows.   
\end{proof}
\newtheorem{thm5}[thm1]{Theorem}  
\begin{thm5}
Let $L(T)$ be the line graph of complete $k$-ary tree $T$ with height $h \geq 2$. Then \[\chi_r(L(T)) = \left\{ 
\begin{array}{l l}
k+1, & \quad \mbox{if $r \leq k$. \textsl{}}\\  
2k+1,   & \quad \mbox {if $ r = \Delta$. \textsl{}}\\ \end{array} \right. \] 
\end{thm5}
\begin{proof}
Let $V(L(T))= \left \{v_1,v_2,\dotsc,v_{e(h)}\right \}$, where $e(h)=\frac{k^{h+1}-1}{k-1}-1$. In $T$  we assume that the root is at level $0$ and for each $l$ ($1 \leq l \leq h$), $v_{e(l-1)+1}$ to $v_{e(l)}$ represent the edges between levels $l-1$ and $l$, numbered from `left' to `right'. It can be  seen that $\Delta(L(T))=2k$ and $\omega(L(T))=k+1$. In the ordering $v_{e(h)}, \dotsc, v_1$ of the vertices of $L(T)$, for each $i$ ($1 \leq i \leq e(h)$), $v_i$ is a simplicial vertex in the subgraph induced by $\{v_i,\dotsc,v_1\}$. Hence the ordering is a p.e.o. and $L(T)$ is chordal. As every chordal graph is perfect,  $\chi(L(T))= \omega(L(T))= k+1$. Since every vertex of $L(T)$ is in a $K_{k+1}$, we also have $\chi_r(L(T))=k+1$ if $r \leq k$. Thus $\chi_r(L(T)) = k+1$, if $r \le k$.  From \cite{Lai06} we know $\chi_r(G) \geq \min \{r, \Delta \}+1$. Taking $G=L(T)$ we have $\chi_r(L(T)) \geq \min \{r, \Delta \}+1=2k+1$ if $r=\Delta$. Similar to the greedy (vertex) coloring, color the vertices in the order $v_1, \dotsc, v_{e(h)}$ by assigning to each vertex the first available color not already used for any of the lower indexed vertices within distance two. In the assumed order, each vertex has at most $\Delta$ lower indexed vertices within distance two; therefore $\chi_\Delta(L(T)) \leq \Delta +1=2k+1$.  Hence  $\chi_r(L(T)) = 2k+1$ if $r = \Delta$.  
\end{proof} 
\noindent \textbf{Definition 3} The wheel graph $W_n$ consists of a $C_{n-1}$ together with a center vertex $s$ that is adjacent to all the $n-1$ vertices in the $C_{n-1}$.
\newtheorem{thm6}[thm1]{Theorem}  
\begin{thm6} 
If $W_n$ denotes the wheel graph, for any $r\geq 3$  \[\chi_r(W_n) = \left\{ 
\begin{array}{l l}
\chi_r(C_{n-1})+1, & \quad \mbox{if $r \leq \chi_r(C_{n-1}).$ \textsl{}}\\  
\min \{r,n-1\}+1, & \quad \mbox{if $ r > \chi_r(C_{n-1}).$\textsl{}}\\ \end{array} \right. \]
\end{thm6}
\begin{proof}
By definition $d(s)=n-1$. For any other $u \; (\neq s) \in V(W_n),\; d(u)=3$.  Since $r \geq 3$, every conditional $(k,r)$-coloring of $W_n$ is also a conditional $(k,r)$-coloring of its induced subgraph $C_{n-1}$. There exists conditional $(k,r)$-coloring of $G$ iff $k \geq \chi_r(G)$. Hence conditional $(\chi_r(C_{n-1}),r)$-coloring of $C_{n-1}$ with a new color to $s$ satisfies (C1) and gives $|c(N_{W_n}(v))| \geq  \min\{r, d(v)\}$, for all $v \in V(W_n)\backslash s$. If  $r \leq \chi_r(C_{n-1})$, then $|c(N_{W_n}(s))| \geq r$. Thus for each $v \in V(W_n)$, $|c(N_{W_n}(v))|\geq \min\{r, d(v)\}$ satisfying (C2) -- in total $\chi_r(C_{n-1})+1$ colors are used. If $r > \chi_r(C_{n-1})$, then $|c(N_{W_n}(s))| < r$, and we need to give new colors to $\min\{r,n-1\}-\chi_r(C_{n-1})$ vertices adjacent to $s$. In total $\chi_r(C_{n-1})+1+\min \{r,n-1\}-\chi_r(C_{n-1})=\min \{r,n-1\}+1$ colors are used. Hence the result.  
\end{proof}  
\noindent \textbf{Definition 4} The $n$-gear $G_n$ consists of a cycle $C_{2n}$ on $2n$ vertices where every other vertex on the cycle is adjacent to a $(2n+1)th$ center vertex labeled $v_0$. The  vertices in the $C_{2n}$ are labeled sequentially $v_1, \ldots,v_{2n}$ such that for  $1 \leq i \leq 2n-1$, $v_i$ is adjacent to $v_{i+1}$, $v_1$ is adjacent to $v_{2n}$,  and every vertex in $V_{oddi}= \{v_i |\; \text{i is odd and}\; 1 \le i \le 2n-1  \}$ is adjacent to $v_0$.
\newtheorem{thm7}[thm1]{Theorem}  
\begin{thm7} 
If $G_n$ is the $n$-gear then for any $n \geq 3$ \[\chi_r(G_n) = \left\{ 
\begin{array}{l l}
4, & \quad \mbox{if $r =2$. \textsl{}}\\  
\chi_2(C_{2n})+1, & \quad \mbox{if $ r =3$.\textsl{}}\\ 
min \{r,\Delta \}+1, & \quad \mbox{if $r \geq 4$. \textsl{}}\\  
\end{array} \right. \]
\end{thm7}
\begin{proof}
Let $k = \chi_r(G_n)$. From ~\cite{Lai06} we know $\chi_r(G) \geq min \{r, \Delta\}+1$. Taking $G=G_n$ we have $\chi_r(G_n) \geq min \{r, \Delta\}+1$. Let $k=\chi_r(G_n)$.\\
\textbf{Case 1:} $r=2:$ We have $k \geq 3$. We assume that $k=3$. Let $c \colon V(G_n) \to \{1,2,3 \}$ be a conditional $(3,2)$-coloring where $c(v_i)=i$ for $i=1,2,3$. Then across $v_1, v_2, v_3$ (C1) must be true and in particular (C2) must hold at $v_2$. To satisfy (C1) at $v_3, c(v_4) \neq 3$. We branch into two cases. If $c(v_4)=1$, then by (C2) at $v_4$ we must have $c(v_5) \notin \{1,3\}$. Therefore we must have $c(v_5)=2$. To satisfy (C1), $c(v_0) \notin \{1,2,3\}$. On the other hand if $c(v_4)=2$ then to satisfy (C2) at $v_4$ we must have $c(v_5) \notin \{2,3\}$. Hence $c(v_5)=1$. To satisfy (C2) at $v_3$ while preserving proper coloring, $c(v_0) \notin \{1,2,3\}$. Therefore $k \geq 4$. To show that $k=4$, it suffices to construct a conditional $(4,2)$-coloring of $G_n$. Define $c \colon V(G_n) \to \{1,2,3,4 \}$ as follows:  \\
\textbf{Case 1.1:} $n \equiv 0 \pmod{3}:$ Set $c(v_0)=4$ and
\[c(v_i) = \left\{ 
\begin{array}{l l}
1, & \quad \mbox{if \ $i \bmod 3= 1$. \textsl{}}\\  
2, & \quad \mbox{if \ $i \bmod 3= 2 $.\textsl{}}\\ 
3, & \quad \mbox{if \ $i \bmod 3= 0$. \textsl{}}\\  
\end{array} \right. \]
\textbf{Case 1.2:} $n \equiv 2 \pmod{3}:$ Modify $c$ in case (1.1) by the assignment $c(v_{2n})=2$.\\
\textbf{Case 1.3:} $n \equiv 1 \pmod{3}:$ Modify $c$ in case (1.1) by the assignments $c(v_{2n-4})=2,c(v_{2n-3})=1,c(v_{2n-2})=3,c(v_{2n-1})=2$ and $c(v_{2n})=3$.\\ 
It can be verified that $c$ is a conditional $(4,2)$-coloring of $G_n$. Thus $k \leq 4$, and hence $\chi_2(G_n)= 4$.\\
\textbf{Case 2:} $r=3:$ We have $k \geq 4$. By (C1) $c(v_i) \neq c(v_0)$ and by (C2) at $v_i$, $c(v_{i+1}) \neq c(v_0)$ for all odd $i$ in the range $1 \leq i \leq 2n-1$. Hence $c(v_i) \neq c(v_0)$ for all $i \; (i \neq 0)$.
Since $d(v_i) \leq 3$ for $1 \leq i \leq 2n$ and $r=3$, a conditional $(\chi_3(G_n),3)$-coloring of $G_n$ gives a conditional $(\chi_2(C_{2n}),2)$-coloring of $C_{2n}$. In turn, a conditional $(\chi_2(C_{2n}),2)$-coloring of $C_{2n}$ with an additional color to $v_0$ gives a conditional $(\chi_3(G_n),3)$-coloring of $G_n$.\\ 
\textbf{Case 3:} $r \geq 4 :$ Let $l= \min \{r,\Delta(G_n) \}$. We know $\chi_r(G_n) \geq l+1 $. Since $v_0$ is the only vertex with $d(v_0) \geq r$ and $\chi_r(C_{2n}) \leq l $, conditional $(l,r)$-coloring of $V(G_n) \setminus \{v_0\}$ such that $|c(V_{oddi})| = l $, with $(l+1)$th color assigned to $v_0$ results in a conditional $(l+1,r)$-coloring of $V(G_n)$. Thus $\chi_r(G_n) \leq l+1 $, and hence $\chi_r(G_n) = \min\{r, \Delta\}+1 $.
\end{proof} 

\end{document}